\documentclass[iop]{emulateapj}
\newcommand{\CVI}{\ion{C}{6}}

\newcommand{\NVI}{\ion{N}{6}}
\newcommand{\NVII}{\ion{N}{7}}

\newcommand{\OVII}{\ion{O}{7}}
\newcommand{\OVIII}{\ion{O}{8}}

\newcommand{\Kalpha}{K$\alpha$}

\newcommand{\Lyalpha}{Ly$\alpha$}
\newcommand{\Lybeta}{Ly$\beta$}
\newcommand{\Lygamma}{Ly$\gamma$}

\newcommand{\Lyepsilon}{Ly$\epsilon$}

\newcommand{\Ne}{\ensuremath{n_{\mathrm{e}}}}

\newcommand{\nO}{\ensuremath{n_{\mathrm{O}}}}
\newcommand{\NH}{\ensuremath{N_{\mathrm{H}}}}

\newcommand{\angstrom}{\ensuremath{\mbox{\AA}}}
\newcommand{\nm}{\ensuremath{\mbox{\nm}}}
\newcommand{\cm}{\ensuremath{\mbox{cm}}}

\newcommand{\km}{\ensuremath{\mbox{km}}}

\newcommand{\pc}{\ensuremath{\mbox{pc}}}

\newcommand{\s}{\ensuremath{\mbox{s}}}
\newcommand{\ks}{\ensuremath{\mbox{ks}}}

\newcommand{\kev}{\ensuremath{\mbox{keV}}}

\newcommand{\erg}{\ensuremath{\mbox{erg}}}

\newcommand{\sr}{\ensuremath{\mbox{sr}}}

\newcommand{\K}{\ensuremath{\mbox{K}}}

\newcommand{\counts}{\ensuremath{\mbox{counts}}}

\newcommand{\parcminsq}{\ensuremath{\mbox{arcmin}^{-2}}}
\newcommand{\pdegsq}{\ensuremath{\mbox{deg}^{-2}}}

\newcommand{\pcmsq}{\ensuremath{\cm^{-2}}}

\newcommand{\pks}{\ensuremath{\ks^{-1}}}
\newcommand{\ps}{\ensuremath{\s^{-1}}}
\newcommand{\psr}{\ensuremath{\sr^{-1}}}

\newcommand{\emismeas}{\ensuremath{\cm^{-6}}\ \pc}

\newcommand{\flux}{\erg\ \pcmsq\ \ps}

\newcommand{\LU}{\ensuremath{\mbox{L.U.}}}
\newcommand{\kmps}{\km\ \ps}

\newcommand{\rassrate}{\counts\ \ps\ \parcminsq}

\newcommand{\ace}{\textit{ACE}}

\newcommand{\iras}{\textit{IRAS}}

\newcommand{\rosat}{\textit{ROSAT}}

\newcommand{\suzaku}{\textit{Suzaku}}

\newcommand{\xmm}{\textit{XMM-Newton}}

\newcommand{\chisq}{\ensuremath{\chi^2}}

\newcommand{\raymondsmith}{\citeauthor{raymond77} (\citeyear{raymond77} and updates)}
\providecommand{\eqref}[1]{equation~(\ref{#1})}

\newcommand{\esas}{\textit{XMM}-ESAS}

\newcommand{\cumbee}{C14}
\newcommand{\Tfg}{\ensuremath{T_\mathrm{fg}}}
\newcommand{\EMfg}{\ensuremath{\mathcal{E}_\mathrm{fg}}}

\newcommand{\Th}{\ensuremath{T_\mathrm{h}}}
\newcommand{\EMh}{\ensuremath{\mathcal{E}_\mathrm{h}}}

\newcommand{\Oplus}[1]{\ensuremath{\mathrm{O}^{#1+}}}

\shorttitle{\textit{XMM-NEWTON} SHADOWING MEASUREMENT OF HALO X-RAY EMISSION}
\shortauthors{HENLEY ET AL.}

\begin{document}

\title{\textit{XMM-Newton} Measurement of the Galactic Halo X-ray Emission using a Compact Shadowing Cloud}
\author{David B. Henley, Robin L. Shelton, Renata S. Cumbee, and Phillip C. Stancil}
\affil{Department of Physics and Astronomy, University of Georgia, Athens, GA 30602; dbh@physast.uga.edu}

\begin{abstract}
Observations of interstellar clouds that cast shadows in the soft X-ray background can be used to
separate the background Galactic halo emission from the local emission due to solar wind charge
exchange (SWCX) and/or the Local Bubble (LB). We present an \xmm\ observation of a shadowing cloud,
G225.60$-$66.40, that is sufficiently compact that the on- and off-shadow spectra can be extracted
from a single field of view (unlike previous shadowing observations of the halo with CCD-resolution
spectrometers, which consisted of separate on- and off-shadow pointings).
We analyzed the spectra using a variety of foreground models: one representing LB emission, and two
representing SWCX emission. We found that the resulting halo model parameters (temperature $\Th
\approx 2 \times 10^6~\K$, emission measure $\EMh \approx 4 \times 10^{-3}~\emismeas$) were not
sensitive to the foreground model used. This is likely due to the relative faintness of the
foreground emission in this observation. However, the data do favor the existence of a foreground.
The halo parameters derived from this observation are in good agreement with those from previous
shadowing observations, and from an \xmm\ survey of the Galactic halo emission. This supports the
conclusion that the latter results are not subject to systematic errors, and can confidently be used
to test models of the halo emission.
\end{abstract}


\section{INTRODUCTION}
\label{sec:Introduction}

An important result from \rosat\ was the discovery of shadows in the soft X-ray background (SXRB),
caused by interstellar clouds partially blocking the distant X-ray emission
\citep{burrows91,snowden91}.  Analysis of such shadows showed that hot, X-ray-emitting plasma exists
in the halo of our Galaxy \citep[e.g.,][]{wang95,kuntz97,snowden00}.  By comparing the X-ray
emission observed toward and to the side of a shadowing cloud, one can separate the hot halo
emission from the foreground emission, attributable to hot gas in the Local Bubble (LB;
\citealt{sanders77,snowden90}), charge exchange (CX) reactions between solar wind ions and neutral H
and He in the heliosphere and the Earth's exosphere
\citep{cravens00,robertson03a,robertson03b,koutroumpa06}, or a combination of the two
\citep{smith14,galeazzi14}. Separating the foreground and halo emission is necessary to test models
for the foreground emission, and for the origin of the hot halo plasma.

More recently, \xmm\ and \suzaku\ observations of shadowing clouds have been used to constrain the
hot halo emission. These satellites' CCD cameras have higher spectral resolution than \rosat's
proportional counter. Such studies obtained halo temperatures and emission measures of $\sim$$2
\times 10^6$~K and $\sim$$\mbox{(3--12)} \times 10^{-3}$~\emismeas, respectively
\citep{galeazzi07,smith07a,gupta09b,lei09}. However, whereas \rosat's large field of view
($\sim$2\degr) meant that a shadowing cloud and the adjacent off-cloud sky could be observed in a
single pointing, \xmm\ and \suzaku's smaller fields of view ($\sim$0\fdg5 and $\sim$0\fdg3,
respectively) required that the above-cited shadowing observations consist of two separate
pointings---one toward and one to the side of the cloud under study. While this strategy would be
fine if the foreground emission were dominated by a constant source, a time-varying source, solar
wind charge exchange (SWCX) emission, is now known to be a major, possibly dominant, contributor to
the foreground emission in the \xmm\ and \suzaku\ band
\citep{koutroumpa07,koutroumpa09a,koutroumpa11}. This SWCX emission is variable on timescales of
$<$1~day to years
\citep{wargelin04,snowden04,fujimoto07,kuntz08a,carter08,henley08a,henley10a,henley12b,carter10,carter11,ezoe11}.
If the foreground SWCX emission varied significantly between the times when the on- and off-shadow
pointings were made, the above shadowing analyses would be inaccurate.

In order to ensure that the foreground contribution to the on- and off-shadow emission would be
identical, we searched the \textit{COBE}/DIRBE-corrected \iras\ dust maps \citep{schlegel98} for
compact interstellar clouds that would potentially cast an X-ray shadow that would fit within a
single \xmm\ field of view. We identified the cloud G225.60$-$66.40 (G225$-$66 in
\citealt{odenwald88}; G225 hereafter) as a viable target (see Figure~\ref{fig:Images}(a)). The
optical depth of this cloud is such that the observed 0.4--1.0~\kev\ surface brightness of the
background emission toward the cloud is $\sim$2/3 of that to the side of the cloud. From simulations
we found that such a cloud would be expected to cast a shadow in a
$\sim$60~\ks\ \xmm\ exposure.\footnote{Another potentially viable target was [RHK93] 9364
  \citep{reach93}, at $l=317\fdg3$, $b=+83\fdg8$. However, the contrast between the on- and
  off-cloud regions within a single \xmm\ field was not expected to be as large as for G225. Also,
  it was not possible to obtain the required exposure from a single pointing.}  Unfortunately, the
distance to this cloud is not known. \citet{odenwald88} assumed a distance of 200~\pc; the clouds in
his sample for which he was able to estimate distances are at similar distances. If G225 is at a
distance of $\sim$200~\pc, it would be beyond the LB.

Here, we present the \xmm\ observation of this cloud, which we used to constrain the Galactic halo
X-ray emission. This is the first measurement of this emission using a single-pointing shadowing
observation with a CCD-resolution spectrometer (\citet{anderson10} carried out similar observations
with \xmm, but their target clouds were at low Galactic latitudes ($b \sim 0\fdg1$), and so measured
the disk X-ray emission rather than the halo emission).  In particular, we tested the sensitivity of
our halo measurement to the assumed foreground model. Recent studies have argued that a combination
of LB and SWCX emission is needed to explain the foreground 1/4~\kev\ emission---\citet{smith14} and
\citet{galeazzi14} attributed $\sim$75\%\ and $\sim$40\%\ of the low-Galactic-latitude
1/4~\kev\ foreground to SWCX, respectively.  At higher energies, \citet{koutroumpa11} attributed
approximately half of the foreground \OVII\ emission in an \xmm\ observation of MBM~12 to
SWCX. However, the relative contributions of LB and SWCX emission to an arbitrary \xmm\ observation
are not known. Therefore, we considered two limiting cases for our foreground model---one in which
LB emission dominates, and one in which SWCX emission dominates (for the latter case, we examined
two different SWCX models).

The remainder of this paper is organized as follows. The observation and data reduction are
described in Section~\ref{sec:Reduction}. The spectral model and the results from the spectral
analysis are presented in Sections~\ref{sec:Method} and \ref{sec:Results}, respectively. We discuss
our results in Section~\ref{sec:Discussion}.

\section{OBSERVATION AND DATA REDUCTION}
\label{sec:Reduction}

\begin{figure*}
\centering
\includegraphics[width=0.425\linewidth]{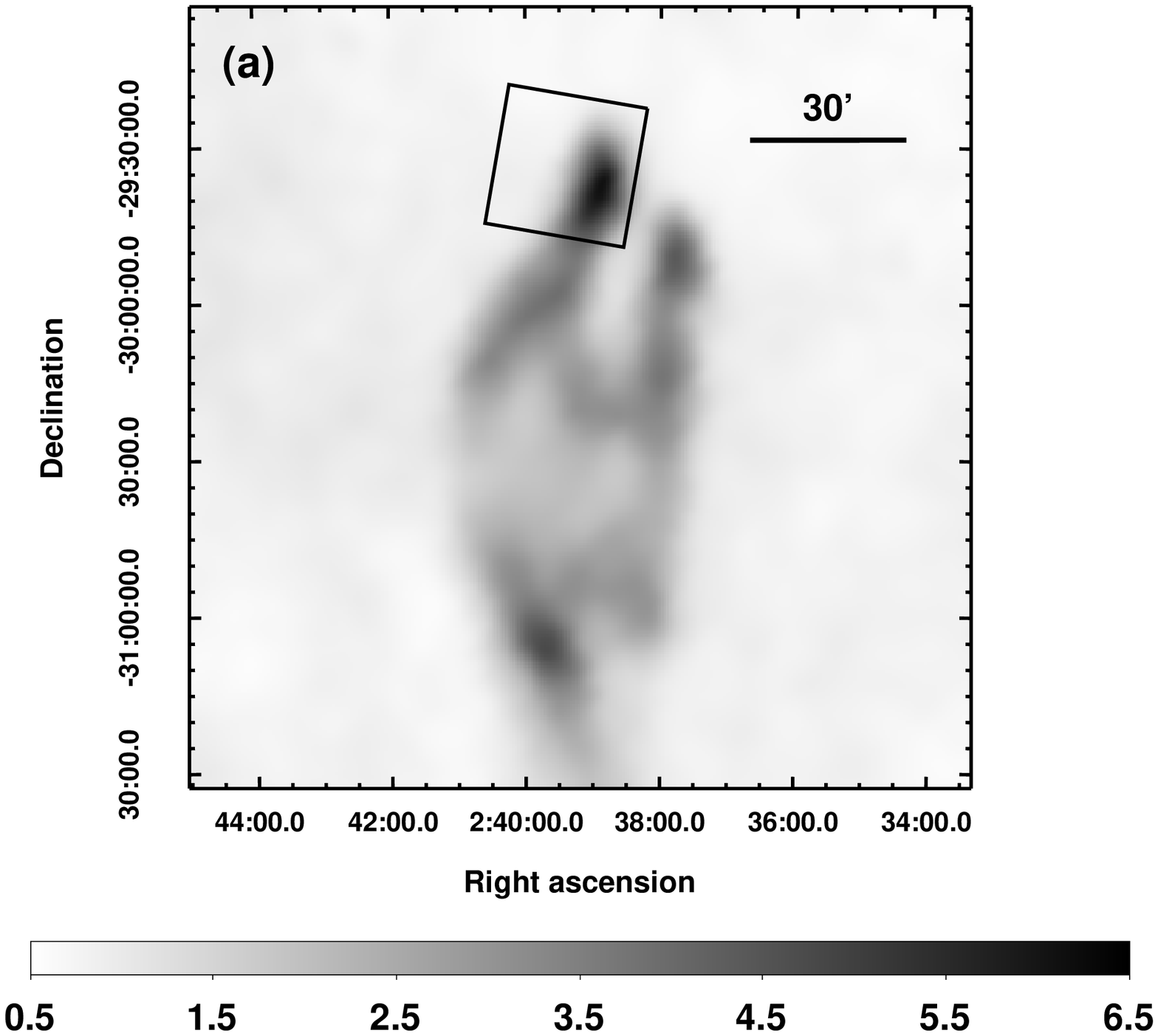} \\
\plottwo{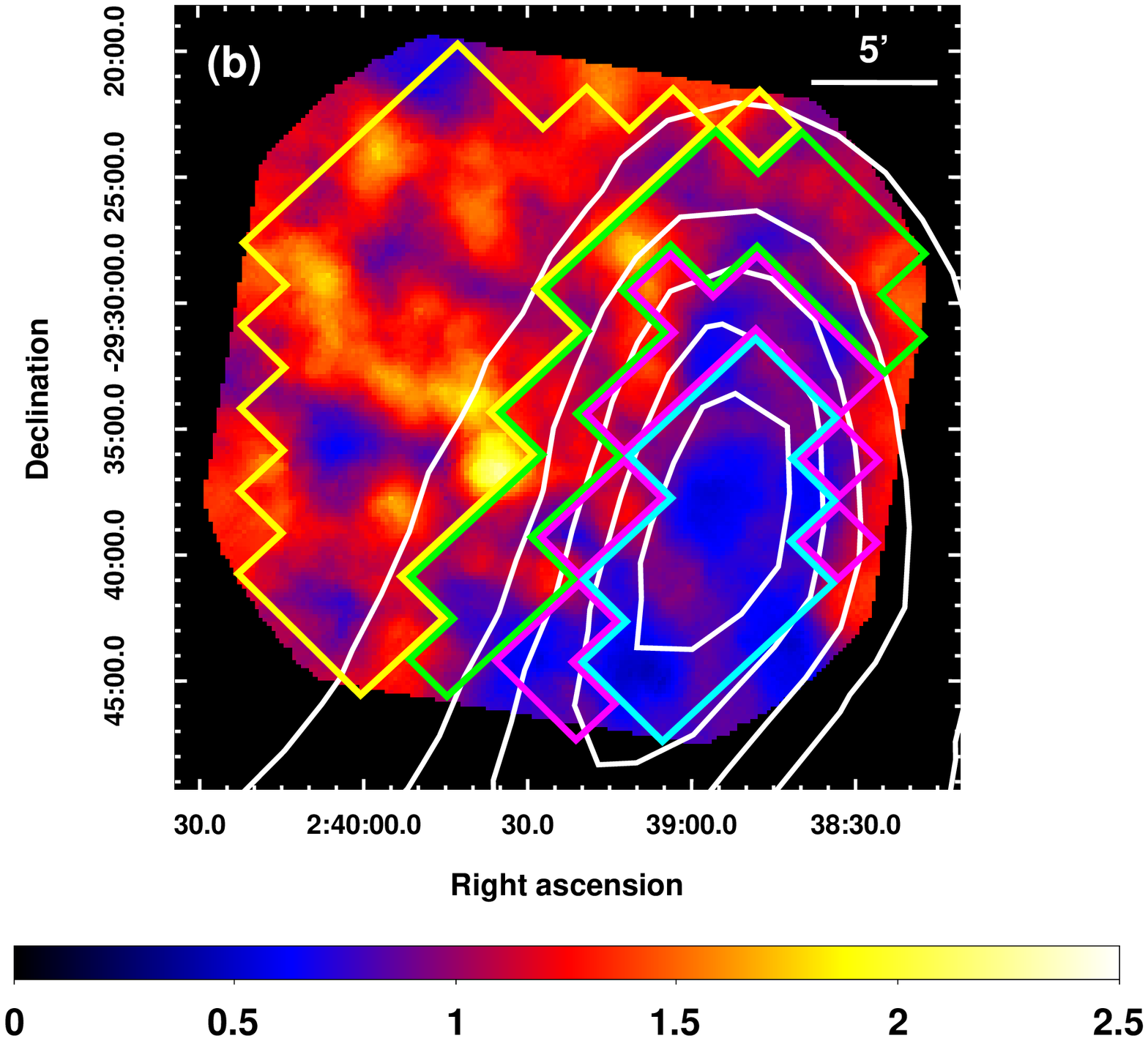}{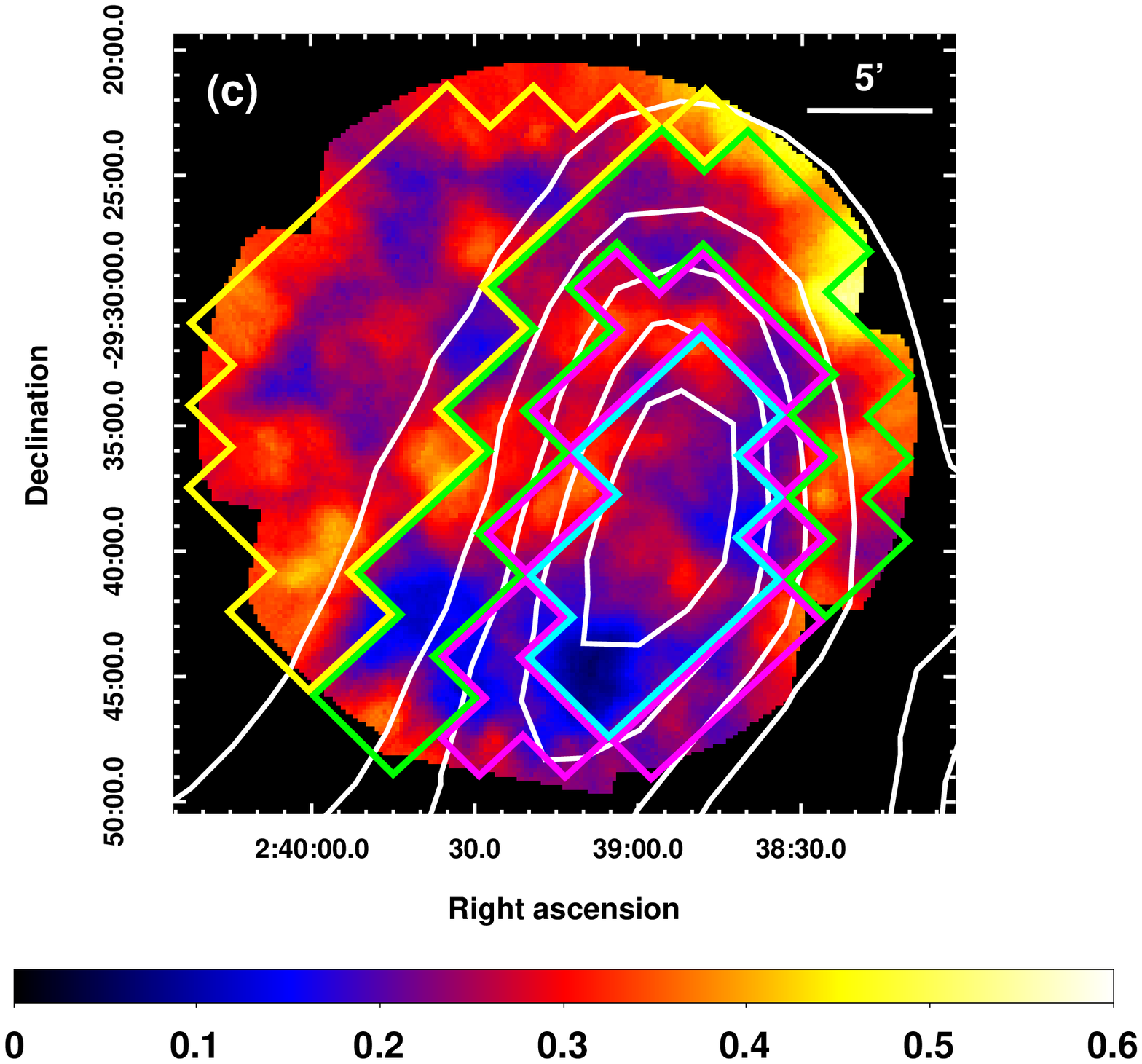}
\caption{(a) \iras\ 100-\micron\ image of G225 \citep{schlegel98}. The gray scale is in MJy
  \psr. The black square indicates the \xmm\ pn field of view.
  (b) QPB- and soft-proton-subtracted, flat-fielded, adaptively smoothed 0.4--1.2~keV \xmm\ pn image
  of G225. The chip gaps and the holes in the data resulting from the source removal have been
  filled using data from neighboring pixels. The color scale is in counts \pks\ \parcminsq. The
  white contours show the \iras\ 100-\micron\ intensity (1--5 MJy \psr\ in one-unit steps). The
  colored polygons indicate the spectral extraction regions (see text for details). Note that the
  polygons used for the spectral extraction follow the pixels in the 100-\micron\ map, whereas
  the contours have been smoothed.
  (c) As in (b), for MOS2.
  \label{fig:Images}}
\end{figure*}

G225 was observed by \xmm\ \citep{jansen01} for 90~ks on 2013 Feb 04--06 (observation ID
0690500101). The pointing direction was $(\alpha,\delta) = (02^\mathrm{h}39^\mathrm{m}20\fs9,
-29\degr35\arcmin51\farcs1)$, or $(l,b) = (225\fdg26,-66\fdg19)$, toward the north-eastern edge of the
cloud (Figure~\ref{fig:Images}(a)).

In our analysis we used data from the EPIC-pn and EPIC-MOS2 cameras (\citealt{struder01,turner01};
note that during the observation only five out of seven MOS1 CCDs were operating---in particular,
the location of one of the inoperative chips meant that $\sim$1/3 of our on-shadow spectral
extraction region [see below] was lost from the MOS1 data). We reduced the data using the
\xmm\ Extended Source Analysis
Software\footnote{http://heasarc.gsfc.nasa.gov/docs/xmm/xmmhp\_xmmesas.html} (\esas;
\citealt{snowden13}), as included in the Science Analysis
System\footnote{http://xmm.esac.esa.int/sas/} (SAS) version 13.5.0. We initially processed the data
using the standard SAS \texttt{epchain} and \texttt{emchain} scripts, and then used the
\esas\ \texttt{pn-filter} and \texttt{mos-filter} scripts to remove periods of soft proton flaring,
during which the count-rate was elevated. After this filtering, 46.6 and 64.1~ks of good time
remained from the pn and MOS2 cameras, respectively.

We used the SAS \texttt{edetect\_chain} script to detect sources with 0.5--2.0~keV fluxes exceeding
$2 \times 10^{-15}~\flux$. Such sources were excluded from the data using circular exclusion
regions. For a given source, the source exclusion radius was equal to the semimajor axis of the
ellipse on which the source count rate per pixel is 0.2 times the local background count rate.  This
radius depends on the source brightness relative to the local background. We estimate that the
0.5--2.0~\kev\ surface brightness of the remaining, unremoved background sources is $(3.0 \pm 0.8)
\times 10^{-12}~\flux\ \pdegsq$ (90\% confidence interval for the whole \xmm\ field). Following
\citet{henley13} and \citet{henley14b}, we based this estimate on the number density of sources with
fluxes of $2.5 \times 10^{-17}$ to $2 \times 10^{-15}~\flux$ \citep{moretti03} and the measurement
of the residual surface brightness after removing sources brighter than $2.5 \times 10^{-17}~\flux$
\citep{hickox06}. The uncertainty estimate takes into account the variance in the number of sources
due to source clustering \citep{peebles80,vikhlinin95} in addition to the Poissonian variance---see
\citet{henley13} for details. The above surface brightness is about twice the typical halo surface
brightness \citep{henley13}. The uncertainty on the surface brightness of the unremoved sources does
not have a statistically significant effect on our measurements (Section~\ref{sec:Results}).

For each camera, we created an image of the 0.4--1.2~keV quiescent particle background (QPB), using
the \esas\ \texttt{pn\_back} and \texttt{mos\_back} programs. These images were constructed using a
database of filter-wheel-closed data, scaled to our observation using data from the unexposed corner
pixels that lie outside the field of view \citep{kuntz08a}. We also used the \esas\ \texttt{proton}
program to create images of the residual soft proton contamination that remains despite the
filtering described above. The parameters for the soft proton models were determined from the
spectral fitting (see Section~\ref{sec:Method}, below). We subtracted the QPB and soft-proton images
from the corresponding 0.4--1.2~keV images extracted from our \xmm\ data, divided these
background-subtracted images by the corresponding exposure maps, and adaptively smoothed the
resulting flat-fielded images (using the \esas\ \texttt{adapt} program). We filled in the chip gaps
and the holes in the data resulting from the source removal using data from neighboring pixels. The
resulting X-ray images of G225 from the pn and MOS2 cameras are shown in Figures~\ref{fig:Images}(b)
and (c), respectively.

In the pn image one can clearly see the shadow cast by the cloud: there is a deficit of counts where
the 100-\micron\ intensity, $I_{100}$, is greatest. However, the shadow is not apparent in the MOS2
image. This difference between the two cameras' images is not an artifact of the particle background
subtraction---the shadow is apparent in the pn image and not the MOS2 image even if we do not
subtract the QPB and the soft proton contamination. Instead, the difference is due to the MOS2
camera's lower sensitivity---for a $\sim$$2\times10^6~\K$ plasma, say, the 0.4--1.2~\kev\ MOS2 count
rate is $\sim$1/5 the pn rate. We used our best-fit spectral model (with an LB foreground component;
see Sections~\ref{subsec:LBForeground} and \ref{sec:Results}, below) to estimate the count rates
expected over the pn and MOS2 fields, taking into account the variation in the absorbing column
density of the cloud and the telescope vignetting. While the pn data are indeed expected to exhibit
a shadow, the resulting MOS2 count rates are too low to produce a noticeable contrast between the
on- and off-shadow regions, given the \xmm\ exposure time.

We extracted X-ray spectra from different regions of the \xmm\ field of view, corresponding to
different absorbing column densities, \NH. These column densities were derived from the
\iras\ $I_{100}$ map \citep{schlegel98}, using the \citet{snowden00} $I_{100}$-to-\NH\ conversion
relation. The spectral extraction regions are shown by the colored polygons in
Figures~\ref{fig:Images}(b) and (c).  These regions outline the $I_{100}$ pixels that correspond to
the following \NH\ ranges: $<2$ (yellow), 2--4 (green), 4--6 (magenta), and
$>6\times10^{20}$~\pcmsq\ (cyan). Note that, because of the different fields of view, the extraction
regions for the MOS2 spectra are slightly different from those for the pn spectra.

From each region we extracted a pn and a MOS2 SXRB spectrum, using the \esas\ \texttt{pn-spectra}
and \texttt{mos-spectra} scripts, respectively, and grouped the resulting spectra such that there
were at least 50 counts per bin. The spectral extraction scripts also calculated the redistribution
matrix file (RMF) and the ancillary response file (ARF) needed for each spectrum, using the SAS
\texttt{rmfgen} and \texttt{arfgen} programs, respectively. For each spectrum, we calculated a
corresponding QPB spectrum using the \esas\ \texttt{pn\_back} and \texttt{mos\_back} programs. As
noted above, the QPB spectra were constructed from a database of filter-wheel-closed data, scaled
using data from the camera pixels outside the field of view. We subtracted from each SXRB spectrum
the corresponding QPB spectrum before carrying out our spectral analysis.

\section{SPECTRAL MODEL DESCRIPTION}
\label{sec:Method}

In order to separate the foreground and halo emission, we used
XSPEC\footnote{http://heasarc.gsfc.nasa.gov/xanadu/xspec/} version 12.8.1l \citep{arnaud96} to fit
an SXRB spectral model simultaneously to the 0.4--5.0~\kev\ spectra extracted from the different
regions of the \xmm\ detectors (we used the spectra from all four regions indicated in
Figures~\ref{fig:Images}(b) and (c)).  Because the pn image exhibits an X-ray shadow whereas the
MOS2 image does not (Figure~\ref{fig:Images}), we investigated fitting to the complete set of pn and
MOS2 spectra and fitting just to the pn spectra.  We assumed \citet{anders89} abundances.

Our SXRB spectral model consisted of components representing emission from the foreground, the
Galactic halo, and the extragalactic background. We also included components representing parts of
the instrumental background that were not removed by the QPB subtraction (see below). As noted in
the Introduction, we experimented with different models for the foreground, described in the
subsections below. In particular, we considered limiting cases in which LB emission
(Section~\ref{subsec:LBForeground}) or SWCX emission (Sections~\ref{subsec:SWCXForeground} and
\ref{subsec:ACXForeground}) dominate the foreground. The details of the other model components are
as follows.

We modeled the Galactic halo emission with a single-temperature ($1T$) APEC thermal plasma model
\citep{smith01a,foster12}, whose temperature and emission measure were free parameters.  We modeled
the extragalactic background using the double broken power-law model described in \citet{smith07a},
but with the overall normalization rescaled so that the 0.5--2.0~\kev\ surface brightness matched
that expected from sources below the source removal flux threshold of $2 \times 10^{-15}~\flux$
\citep{henley13,henley14b}; as noted in Section~\ref{sec:Reduction}, this surface brightness is $3.0
\times 10^{-12}~\flux\ \pdegsq$. These components were subject to absorption, modeled using the
XSPEC \texttt{phabs} model \citep{balucinska92,yan98}. The absorbing column density, \NH, was
different for each spectral extraction region, and was calculated from the average value of
$I_{100}$ in each region \citep{schlegel98}, using the conversion relation from
\citet{snowden00}. These column densities were 1.47 (1.50), 2.78 (2.82), 4.94 (4.96), and
$7.00\ (7.00) \times 10^{20}~\pcmsq$ for the yellow, green, magenta, and cyan regions in
Figure~\ref{fig:Images}(b) (Figure~\ref{fig:Images}(c)), respectively. At the energy of the
\OVII\ line, the optical depth in the highest-\NH\ region is 0.66, meaning that the halo
\OVII\ emission is attenuated by 48\%. In the lowest-\NH\ region, the halo \OVII\ emission is
attenuated by 13\%.

In addition to the above SXRB components, we added Gaussians at $\sim$1.49 and $\sim$1.75~\kev\ to
model the Al and Si instrumental fluorescence lines, respectively (note that spectra from the pn
detector do not exhibit the Si line). These lines are not included in the QPB spectra calculated
using \esas, and hence were not removed by the QPB subtraction. The parameters of these lines were
independent for each individual spectrum. In order to model any residual soft proton contamination
that remained in the spectra despite the filtering described in Section~\ref{sec:Reduction}, we
added a power-law that was not folded through the instrumental response
\citep{kuntz08a,snowden13}. For each detector used (pn or MOS2), the spectral index of this
component was the same for all four spectra, and the normalizations were tied together according to
the relative scaling given by the \esas\ \texttt{proton\_scale} program. The best-fit parameters of
this soft proton component were used to create the soft proton images mentioned in
Section~\ref{sec:Reduction}, which were used in the creation of Figures~\ref{fig:Images}(b) and (c).

\subsection{Foreground Model 1: Local Bubble (LB)}
\label{subsec:LBForeground}

We initially modeled the foreground emission with a $1T$ APEC thermal plasma model that was not
subject to any absorption.  The temperature and emission measure of this component were free
parameters. Physically, this model represents emission from a hot plasma, like that thought to be in
the LB. Although SWCX is now known to be a major, possibly dominant, source of the foreground
emission in the \xmm\ band \citep{koutroumpa07,koutroumpa09a,koutroumpa11}, such a thermal plasma
model has been found to adequately model the foreground emission in CCD-resolution SXRB spectra
\citep[e.g.,][]{galeazzi07,henley08a,gupta09b}. Note that we assumed that the LB emission originates
entirely in front of the cloud.

\subsection{Foreground Model 2: C14-SWCX}
\label{subsec:SWCXForeground}

While using a thermal plasma model for the foreground emission appears to provide adequate fits to
CCD-resolution SXRB spectra, it is possible that the true shape of the foreground spectrum, likely
dominated by SWCX emission, is different from that expected from a hot plasma.  If this is the case,
then a thermal plasma model for the foreground could lead to biases in the best-fit halo
parameters. Therefore, in an attempt to avoid such biases, we modified our original SXRB model so
that the foreground component was composed of CX emission lines. For this model, we use CX line
ratio data from \citet[hereafter \cumbee; in that paper, we applied our CX data to a
  \suzaku\ observation of the Cygnus Loop, the spectrum of which is different from that of the SWCX
  emission]{cumbee14}. We refer to this new foreground model, which is more physically justified
than a thermal plasma model, as the C14-SWCX model.

This foreground SWCX model consisted of \CVI\ \Lyalpha--$\epsilon$, \OVII\ \Kalpha--$\epsilon$, and
\OVIII\ \Lyalpha--$\epsilon$ emission lines. For the \OVII\ \Kalpha\ feature, we modeled the
forbidden, intercombination, and resonance lines individually. The overall normalization of the
emission from each ion was independent (i.e., we did not constrain the ion ratios \textit{a
  priori}). For each ion, we tied together the lines' normalizations using the relative intensities
from the CX model described in \cumbee.\footnote{Note that the model used here includes
  \CVI\ \Lyepsilon, which was not included in \cumbee. The \CVI\ \Lyepsilon/\Lyalpha\ ratio that we
  used is 0.0012 (R.~S. Cumbee \& P.~C. Stancil, 2014, private communication).} These CX line ratios
were calculated for a collision energy of 1~\kev\ u$^{-1}$ (438~\kmps; cf.\ a typical speed for the
slow solar wind is 400~\kmps; e.g., \citealt{smith03}). Note that the \cumbee\ CX data are for ions
interacting with H. However, as He is an order of magnitude less abundant than H, and CX
cross-sections involving He are typically smaller than those involving H
\citep[e.g.,][Table~1]{koutroumpa06}, neglecting interactions between solar wind ions and He should
not adversely affect our results. Note also that, because of the relatively poor spectral resolution
of the \xmm\ detectors at low energies, we did not include lines from \NVI\ or \NVII\ in the
C14-SWCX model (these ions' \Kalpha\ lines lie between those of \CVI\ and \OVII).

\citet{carter10} and \citet{ezoe11} used a similar CX model (based on data from
\citealt{bodewits07}) in their analyses of SWCX enhancements observed during an \xmm\ and a
\suzaku\ observation, respectively. However, we are unaware of such a model having previously been
applied to a shadowing observation.

\subsection{Foreground Model 3: ACX-SWCX}
\label{subsec:ACXForeground}

Our third and final foreground model used the AtomDB Charge Exchange code \citep[ACX;][]{smith14},
and is referred to here as ACX-SWCX.  For each ion receiving an electron via CX, the ACX model uses
analytic expressions to calculate the most-probable $n$ shell and the distribution of orbital
angular momenta, $l$, for the captured electron (see \citealt{smith14} for details). This model then
calculates the spectrum produced as the electron radiatively cascades to the ground state (mainly
using data from AtomDB 2.0.2; \citealt{foster12}). The relative strengths of the lines from
different ions of the same element are determined from the ionization balance of the input ion
population, which is controlled by the model's temperature parameter, assuming that the relative ion
populations are in collisional ionization equilibrium (CIE). The relative strengths of lines from
different elements, meanwhile, are governed by the assumed abundances \citep{anders89}.

For our purposes, we set the ACX model's \texttt{swcx} and \texttt{model} flags to 1 and 8,
respectively \citep{smith14a}. The former setting means that each ion undergoes a single CX reaction
on the line of sight, and is the appropriate setting for studying CX in the context of the diffuse
SXRB. The latter setting means that, if the most-probable $n$ shell for electron capture is not an
integer, the captured electrons are distributed between the two nearest $n$ shells. This setting
also means that the ``Separable'' distribution \citep[Equation~(4)]{smith14} is used for the $l$
distribution.

\section{SPECTRAL ANALYSIS RESULTS}
\label{sec:Results}

\begin{figure}
\centering
\plotone{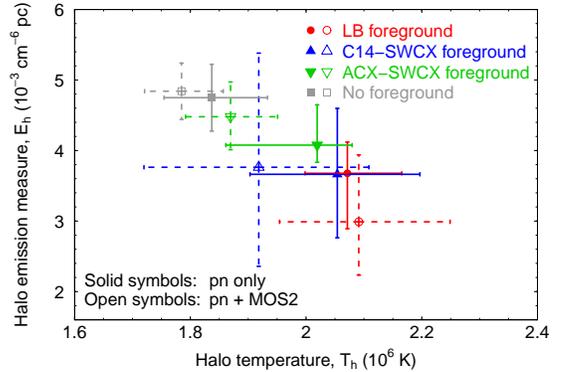}
\caption{Halo temperatures and emission measures obtained using the various foreground models,
  indicated by color (see key). Solid symbols and error bars (open symbols and dashed error bars)
  indicate the results obtained from just the pn data (from the pn and MOS2 data jointly).
  \label{fig:HaloResults}}
\end{figure}

\begin{figure*}
\centering
\plottwo{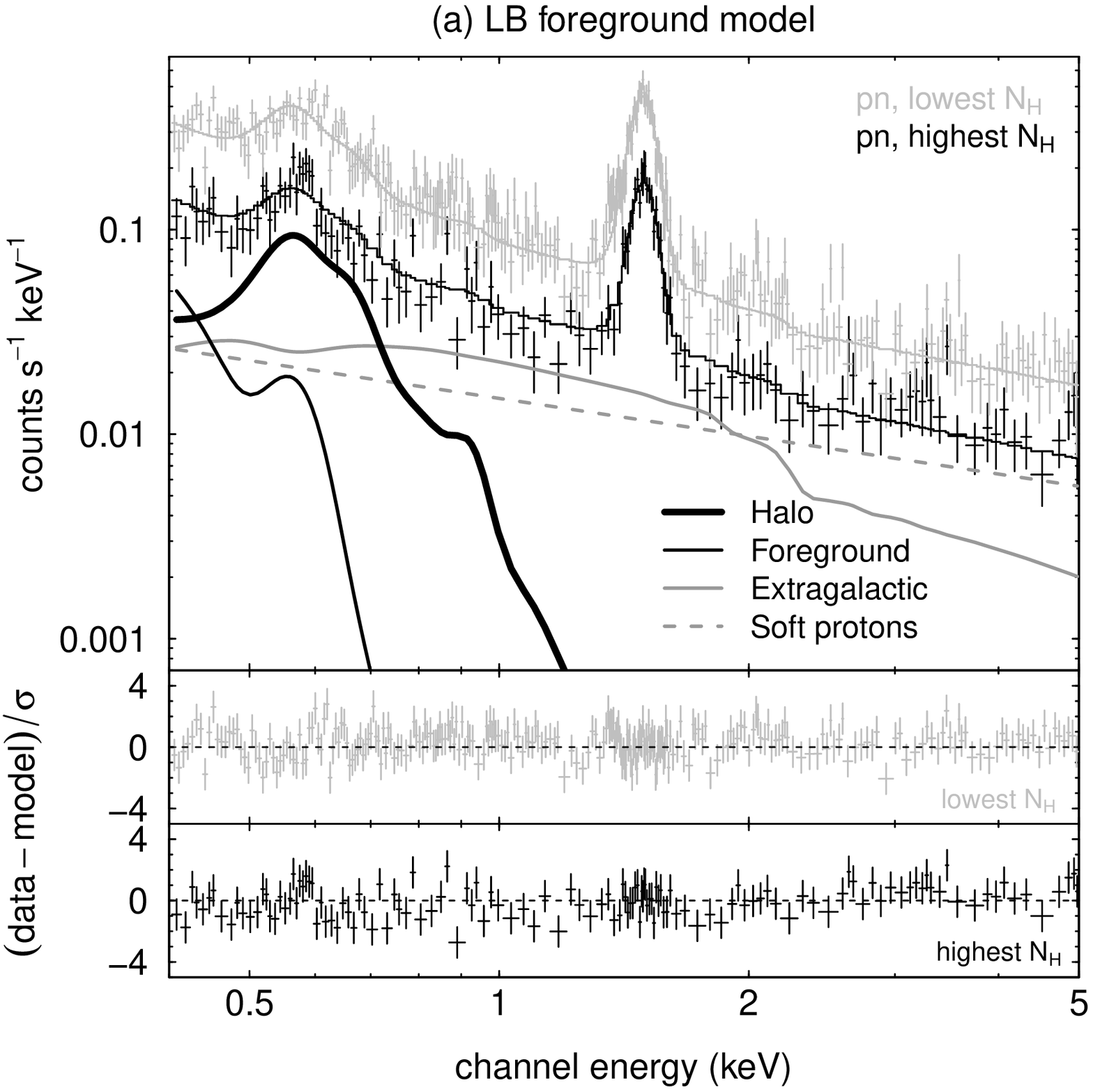}{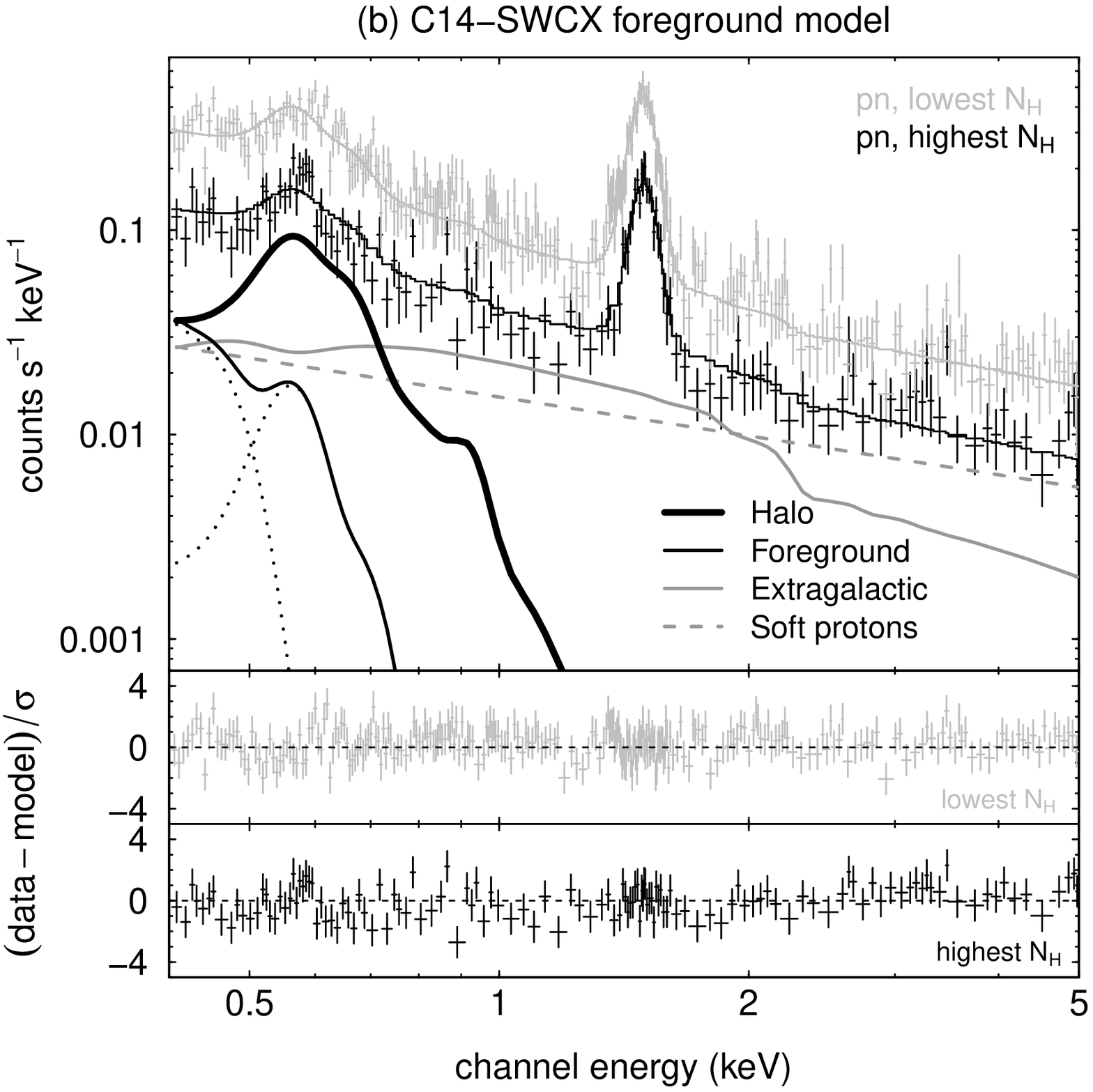}
\plottwo{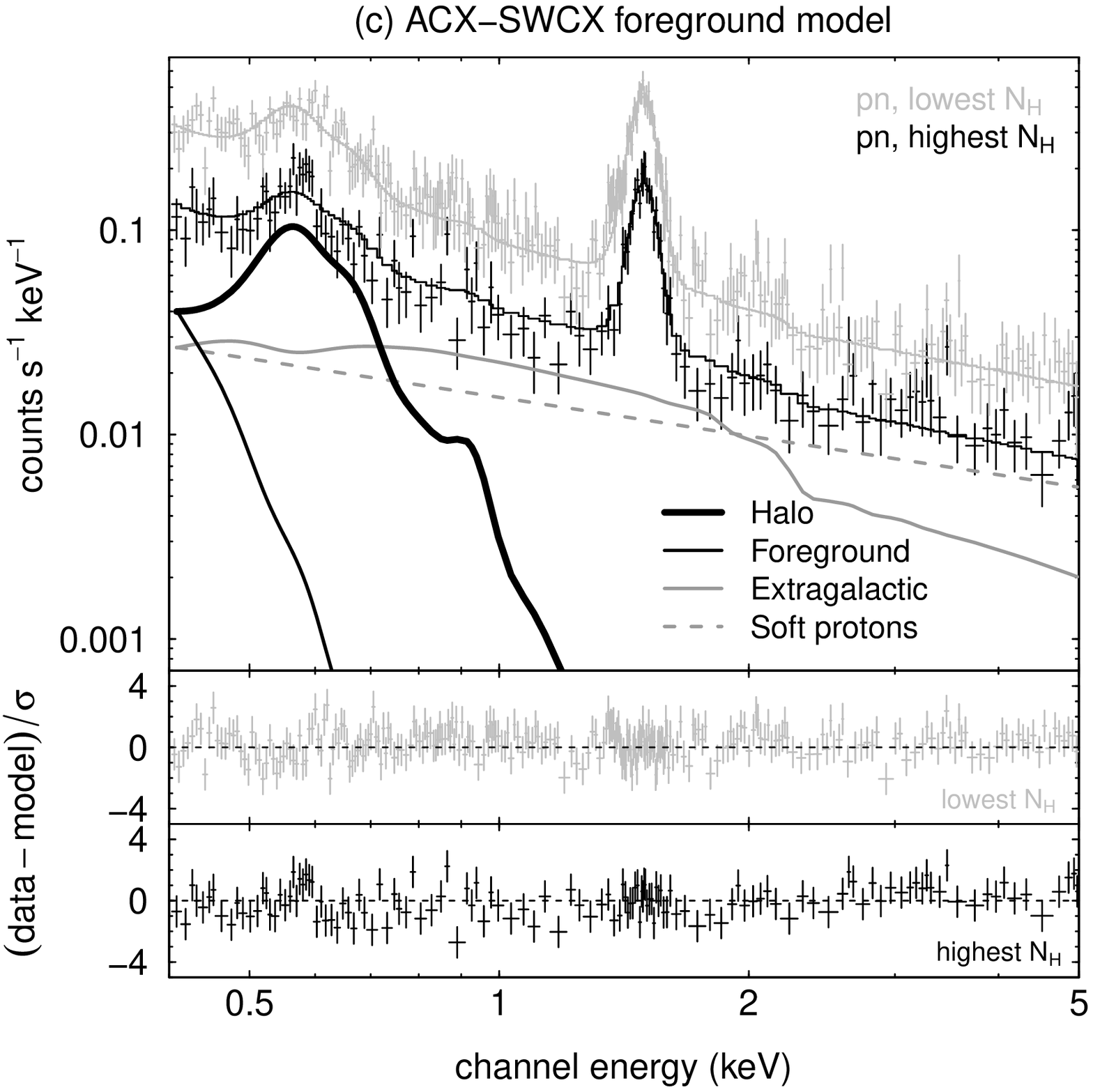}{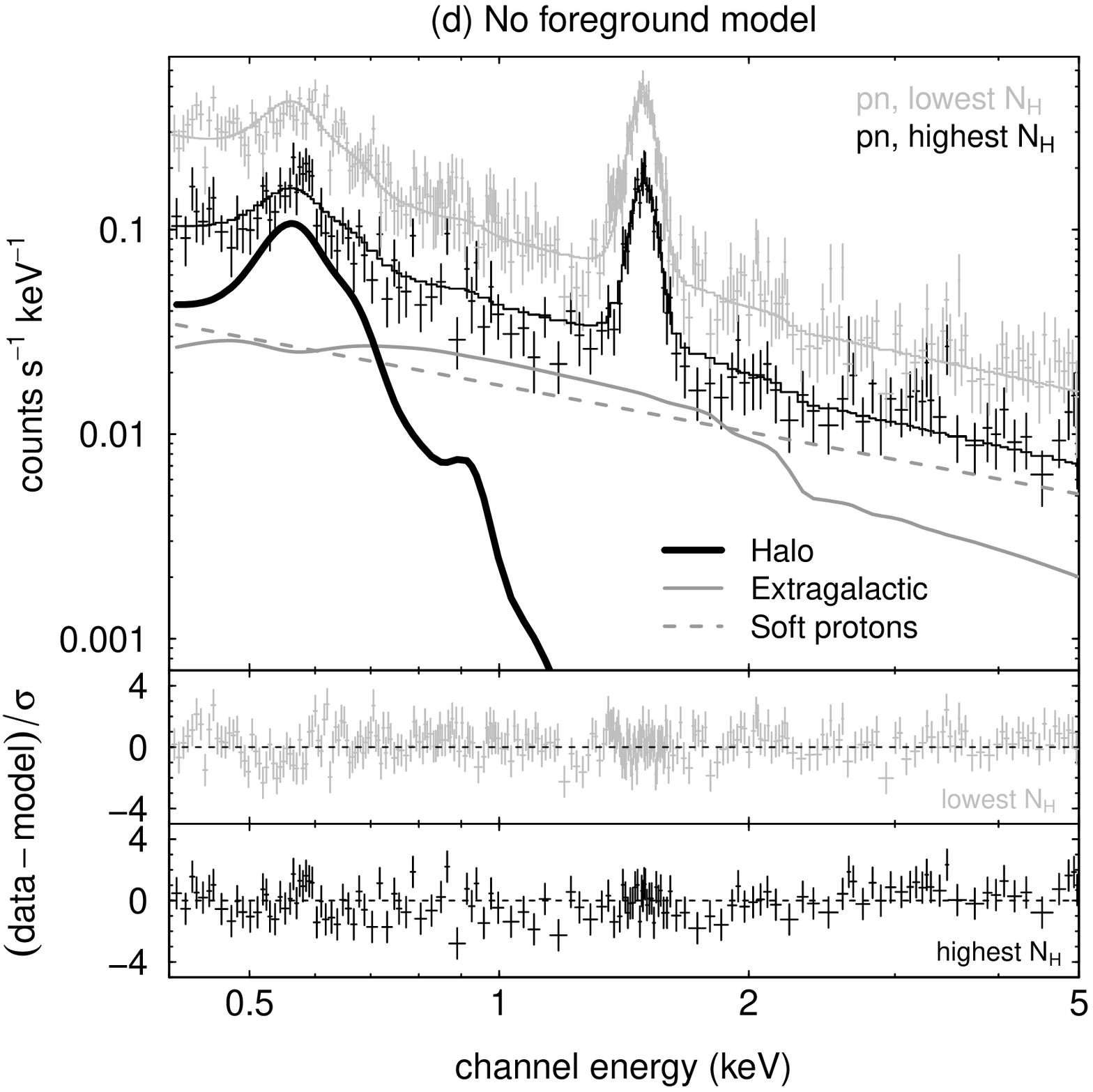}
\caption{\xmm\ pn spectra from the regions of the G225 field with the lowest and highest values of
  \NH\ (gray and black data points in the above plots, corresponding to the yellow and cyan regions
  in Figure~\ref{fig:Images}(b), respectively), with the best-fit spectral models from the fits just
  to the pn data.
  For plotting purposes only, the data have been regrouped such that each bin has a signal-to-noise
  ratio of at least 3.
  Plots (a), (b), and (c) show the best-fit models obtained with the LB
  (Section~\ref{subsec:LBForeground}), C14-SWCX (Section~\ref{subsec:SWCXForeground}), and ACX-SWCX
  (Section~\ref{subsec:ACXForeground}) foreground models, respectively. Plot (d) shows the fit with
  no foreground component in the spectral model.
  For the spectrum from the highest-\NH\ region, we also plot individual model components (see key;
  note that we do not plot the component representing the instrumental Al line). For the SWCX
  foreground model, the dotted lines show the contributions to the foreground from \CVI\ and \OVII\
  (from left to right; the best-fit foreground \OVIII\ intensity is zero).
  \label{fig:Spectra}}
\end{figure*}

The spectral fit results are shown in Table~\ref{tab:Results}, for the LB
(Section~\ref{subsec:LBForeground}), C14-SWCX (Section~\ref{subsec:SWCXForeground}), and ACX-SWCX
(Section~\ref{subsec:ACXForeground}) foreground models. In addition, we show results obtained with
no foreground component in the spectral model (``None''). The upper half of the table shows the
results obtained by fitting simultaneously to the pn and MOS2 spectra, while the lower half shows
the results obtained by fitting just to the pn spectra. The best-fit foreground model parameters are
in columns~2 and 3 for the LB and ACX-SWCX foreground models, and in columns~4--6 for the C14-SWCX
foreground model. For all models, the best-fit halo temperature, \Th, and emission measure, \EMh,
are in columns 7 and 8, respectively.  Figure~\ref{fig:HaloResults} compares the halo temperatures
and emission measures obtained using the various foreground models. Figure~\ref{fig:Spectra} shows
the pn spectra from the regions with the lowest and highest values of \NH\ (yellow and cyan regions
in Figure~\ref{fig:Images}(b), respectively), along with the best-fit models obtained using each of
the three foreground models, and using no foreground model. In general, the fits shown are
reasonably good, and the fits to the spectra that aren't shown are of similar quality.

{
\tabletypesize{\scriptsize}
\setlength{\tabcolsep}{1.5pt}
\begin{deluxetable*}{lrr@{,}lrr@{,}lrr@{,}lrr@{,}lrr@{,}lcrr@{,}lrr@{,}lr@{/}l}
\tablecaption{Spectral Fit Results\label{tab:Results}}
\tablehead{
                     & \multicolumn{15}{c}{Foreground}                                                                                                                                                                                                                                 && \multicolumn{6}{c}{Halo} \\
\cline{2-16} \cline{18-23}
\colhead{Foreground} & \multicolumn{3}{c}{\Tfg}        & \multicolumn{3}{c}{Normalization\tablenotemark{a}} & \multicolumn{3}{c}{$I(\mbox{\CVI})$\tablenotemark{b}} & \multicolumn{3}{c}{$I(\mbox{\OVII})$\tablenotemark{c}} & \multicolumn{3}{c}{$I(\mbox{\OVIII})$\tablenotemark{d}} && \multicolumn{3}{c}{\Th}         & \multicolumn{3}{c}{\EMh}                  & \chisq & dof \\
\colhead{model}      & \multicolumn{3}{c}{($10^6~\K$)} & \multicolumn{3}{c}{}                               & \multicolumn{3}{c}{(\LU)}                             & \multicolumn{3}{c}{(\LU)}                              & \multicolumn{3}{c}{(\LU)}                               && \multicolumn{3}{c}{($10^6~\K$)} & \multicolumn{3}{c}{($10^{-3}~\emismeas$)}                \\
\colhead{(1)}        & \multicolumn{3}{c}{(2)}         & \multicolumn{3}{c}{(3)}                            & \multicolumn{3}{c}{(4)}                               & \multicolumn{3}{c}{(5)}                                & \multicolumn{3}{c}{(6)}                                 && \multicolumn{3}{c}{(7)}         & \multicolumn{3}{c}{(8)}                   & \multicolumn{2}{c}{(9)}
}
\startdata
\multicolumn{10}{l}{Joint fits to pn and MOS2 data:} \\
      LB &  1.02 &  (0.64 &  1.22)               &   7.81 &  (4.16 & 217.37)              & \multicolumn{3}{c}{\nodata} & \multicolumn{3}{c}{\nodata} & \multicolumn{3}{c}{\nodata} &&  2.09 &  (1.95 &  2.25)     &  2.99 &  (2.23 &  3.94)     & 2154.90 & 2032 \\
C14-SWCX & \multicolumn{3}{c}{\nodata}           & \multicolumn{3}{c}{\nodata}            &  1.74 &  (0.48 &  3.17)     &  0.61 &  (0.23 &  2.03)     &  0.01 &  (0.00 &  0.55)     &&  1.92 &  (1.72 &  2.11)     &  3.76 &  (2.36 &  5.38)     & 2164.40 & 2031 \\
ACX-SWCX &  0.63 & \multicolumn{2}{c}{($<$0.78)} &   1.36 & \multicolumn{2}{c}{($>$0.35)} & \multicolumn{3}{c}{\nodata} & \multicolumn{3}{c}{\nodata} & \multicolumn{3}{c}{\nodata} &&  1.87 &  (1.79 &  1.95)     &  4.48 &  (4.01 &  4.97)     & 2160.83 & 2032 \\
    None & \multicolumn{3}{c}{\nodata}           & \multicolumn{3}{c}{\nodata}            & \multicolumn{3}{c}{\nodata} & \multicolumn{3}{c}{\nodata} & \multicolumn{3}{c}{\nodata} &&  1.78 &  (1.72 &  1.86)     &  4.84 &  (4.44 &  5.24)     & 2171.71 & 2034 \\ \\
\multicolumn{10}{l}{Fits to pn data only:} \\
      LB &  0.63 & \multicolumn{2}{c}{($<$1.02)} & 123.35 & \multicolumn{2}{c}{($>$6.97)} & \multicolumn{3}{c}{\nodata} & \multicolumn{3}{c}{\nodata} & \multicolumn{3}{c}{\nodata} &&  2.07 &  (2.00 &  2.17)     &  3.68 &  (2.89 &  4.12)     & 1296.92 & 1265 \\
C14-SWCX & \multicolumn{3}{c}{\nodata}           & \multicolumn{3}{c}{\nodata}            &  2.69 &  (1.93 &  3.44)     &  0.57 &  (0.00 &  1.67)     &  0.00 &  (0.00 &  0.27)     &&  2.05 &  (1.90 &  2.20)     &  3.66 &  (2.76 &  4.60)     & 1302.90 & 1264 \\
ACX-SWCX &  0.75 &  (0.53 &  1.28)               &   0.29 &  (0.01 &  12.23)              & \multicolumn{3}{c}{\nodata} & \multicolumn{3}{c}{\nodata} & \multicolumn{3}{c}{\nodata} &&  2.02 &  (1.86 &  2.08)     &  4.08 &  (3.83 &  4.65)     & 1301.56 & 1265 \\
    None & \multicolumn{3}{c}{\nodata}           & \multicolumn{3}{c}{\nodata}            & \multicolumn{3}{c}{\nodata} & \multicolumn{3}{c}{\nodata} & \multicolumn{3}{c}{\nodata} &&  1.84 &  (1.75 &  1.93)     &  4.75 &  (4.28 &  5.22)     & 1317.11 & 1267
\enddata
\tablecomments{Values in parentheses are the 90\%\ confidence intervals.}
\tablenotetext{a}{For the LB foreground model, this is the foreground emission measure, \EMfg, in units of $10^{-3}~\emismeas$.
  For the ACX-SWCX foreground model, this is the normalization of the foreground component, in units of $10^{-6}~\parcminsq$.}
\tablenotetext{b}{Foreground \CVI\ \Lyalpha\ intensity. As this line ($E = 0.3673~\kev$) is below the
  \xmm\ band used here, this intensity is not constrained directly, but is instead constrained by
  the higher-energy Lyman lines via the C14 CX line ratios.}
\tablenotetext{c}{Foreground \OVII\ \Kalpha\ intensity. We have summed the intensities of the resonance,
  intercombination, and forbidden lines.}
\tablenotetext{d}{Foreground \OVIII\ \Lyalpha\ intensity.}
\end{deluxetable*}
}

\begin{figure}
\centering
\plotone{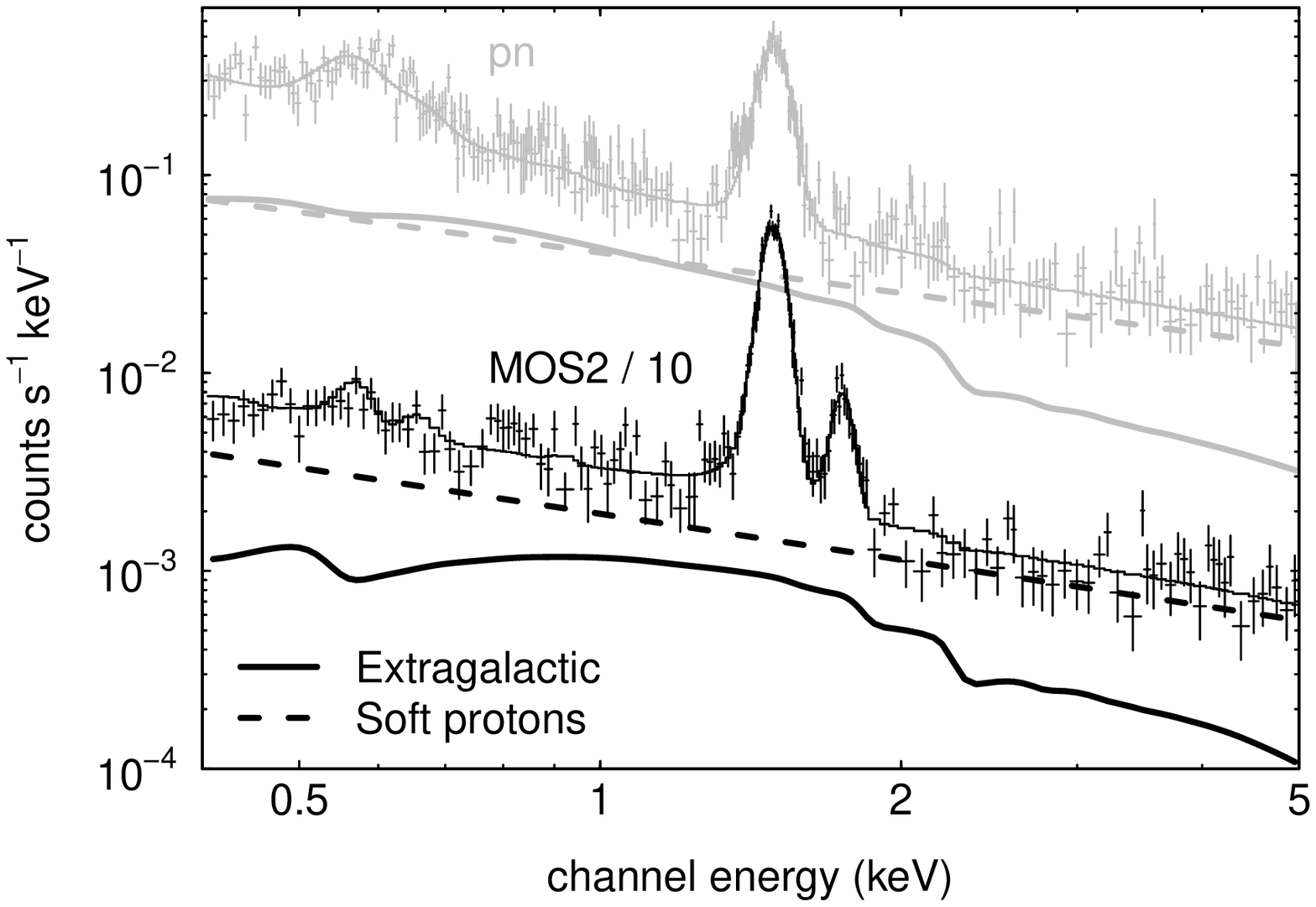}
\caption{\xmm\ pn (gray) and MOS2 (black) spectra from the region of the G225 field with the lowest
  value of \NH, with the best-fit spectral model obtained with the LB foreground model (thin solid
  lines). The MOS2 data have been shifted down by a factor of 10. For each spectrum, we also plot
  the extragalactic and soft proton components of the model (thick solid and dashed lines,
  respectively; the other model components are not plotted). Note that, in the MOS2 spectrum, the
  soft proton component is brighter relative to the extragalactic component than in the pn spectrum,
  indicating more severe soft proton contamination.
  \label{fig:pn+MOS2}}
\end{figure}

Overall, the pn data result in tighter constraints on the halo parameters when used on their own
than when combined with the MOS2 data. The average widths of the 90\%\ confidence intervals on the
halo temperature and emission measure are $0.21 \times 10^6~\K$ and $1.2 \times 10^{-3}~\emismeas$,
respectively, from the pn-only fits, compared with $0.24 \times 10^6~\K$ and $1.6 \times
10^{-3}~\emismeas$, respectively, from the joint pn + MOS2 fits. This difference may be due to the
fact that the soft proton contamination in the MOS2 spectra is more severe than in the pn spectra
(Figure~\ref{fig:pn+MOS2}). Because the pn spectra result in tighter constraints overall on the halo
parameters, in the following we shall concentrate on the results obtained from the pn-only fits.

The results in Table~\ref{tab:Results} were obtained assuming that the 0.5--2.0~\kev\ surface
brightness of the extragalactic background is equal to that expected from sources below the source
removal flux threshold, $3.0 \times 10^{-12}~\flux\ \pdegsq$ (Section~\ref{sec:Method}). The
uncertainty on this expected surface brightness is $\pm 0.8 \times 10^{-12}~\flux\ \pdegsq$
(Section~\ref{sec:Reduction}). We found that varying the surface brightness of the extragalactic
background model within this uncertainty did not have a statistically significant effect on our
best-fit model parameters. This was mainly because, if we adjusted the normalization of the
extragalactic model, the normalization of the soft proton contamination model adjusted itself to
compensate, leaving the other model components not significantly affected.

For the LB foreground model, while the best-fit foreground temperature, \Tfg, is rather low, within
the uncertainty it is consistent with the range of values found from previous shadowing studies
($\Tfg \sim (\mbox{0.8--1.2}) \times 10^6~\K$;
\citealt{snowden00,smith07a,galeazzi07,henley07,henley08a,lei09,gupta09b}). Because this foreground
model is relatively faint within the \xmm\ band (most of the emission would be emitted below
0.4~\kev), its emission measure, \EMfg, is poorly constrained. However, it too is consistent (within
its uncertainty) with the results from previous shadowing studies.

Although the physical nature of the C14-SWCX foreground model is quite different from that of the LB
foreground model, for this particular shadowing observation these two models yield best-fit
foreground spectra that are similar in shape in the \xmm\ bandpass (compare
Figures~\ref{fig:Spectra}(a) and (b)). As a result, the best-fit halo parameters from these two
models are very similar. However, the halo parameters are less well constrained when we use the
C14-SWCX foreground model. This is because, in this model, the foreground \CVI, \OVII, and
\OVIII\ intensities are completely independent, whereas in the LB model they are controlled by the
foreground temperature. This means that there is more freedom in the shape of the foreground
spectrum, and as a result more freedom in the shape of the halo spectrum, and hence in the halo
temperature. Note that the C14-SWCX foreground model yields a higher \chisq\ than the LB foreground
model, despite having one more free parameter.

The ACX-SWCX foreground model yields a much softer best-fit foreground spectrum than the other
foreground models. Since this foreground model produces very little \OVII\ emission, the halo
component must produce relatively more \OVII, and as a result this foreground model yields a
slightly lower halo temperature. However, the difference is only a $\mathrm{few} \times 10^4~\K$,
and is not significant given the error bars.

\begin{figure}
\plotone{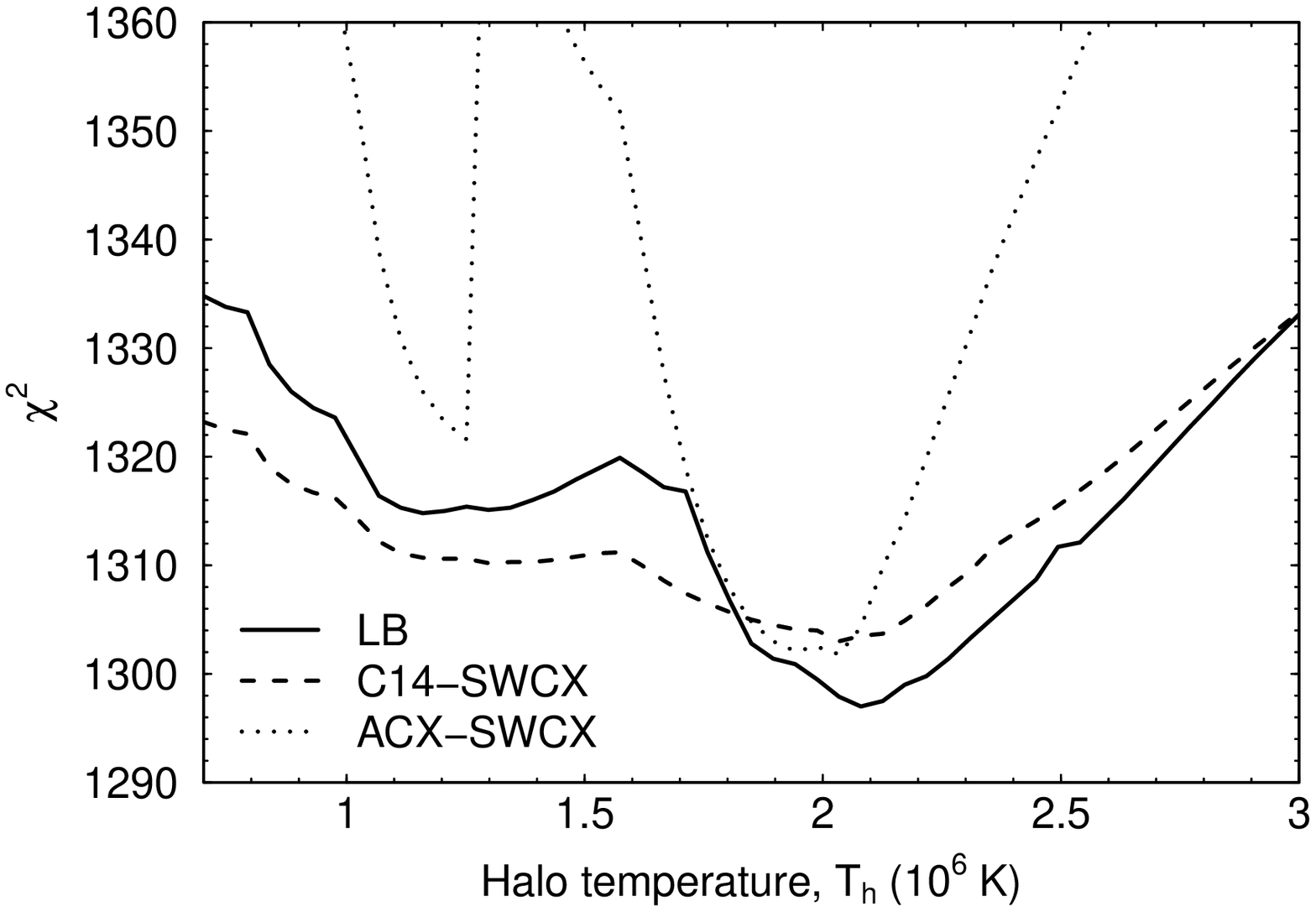}
\caption{\chisq\ as a function of halo temperature for each of the three foreground models that
we studied (solid line: LB model; dashed line: C14-SWCX model; dotted line: ACX-SWCX model).
\label{fig:chisq}}
\end{figure}

Figure~\ref{fig:chisq} shows \chisq\ as a function of halo temperature for each of the foreground
models that we studied. In addition to the best-fit \chisq\ minimum at $\Th \approx 2 \times
10^6~\K$, each curve also exhibits a local minimum at $\Th \approx 1.2 \times 10^6~\K$. At these
local minima, the foreground models are harder than in the best fits, to compensate for the softness
of the cooler halo models. This means that there is some degeneracy between the hardnesses of the
foreground and halo components. However, the differences in \chisq\ between the minima at the lower
and higher halo temperatures are 17.8, 7.2, and 19.8 for the LB, C14-SWCX, and ACX-SWCX foreground
models, respectively, meaning that the lower halo temperature is excluded at the $>$99\%\ level
($\Delta \chisq = 6.63$ for a single interesting parameter; \citealt{lampton76}). To put this
another way, the observed \xmm\ spectra require a soft line-emission component and a hard
line-emission component (with temperatures of $\la$$1.2 \times 10^6$ and $\sim$$(\mbox{2.0--2.5})
\times 10^6~\K$, respectively, for models with a temperature parameter). Figure~\ref{fig:chisq}
shows that models in which the softer component is in the foreground and the harder component is in
the halo (i.e., our best-fitting models, with $\Th \approx 2 \times 10^6~\K$) are strongly preferred
over models in which these two components are switched.

Because the ACX-SWCX foreground model yields similar halo parameters to the other foreground models,
despite the foreground spectrum being much softer, and because omitting the foreground component
altogether still yields an acceptable fit (reduced $\chisq = 1.04$), one could ask if it is
necessary to include a foreground component in the spectral model. To address this question, we used
the Akaike Information Criterion \citep[AIC; e.g.,][]{takeuchi00,liddle07} to determine the relative
quality of the models. The AIC is given by
\begin{equation}
  \mathrm{AIC} = -2 \ln \mathcal{L}_\mathrm{max} + 2k,
  \label{eq:AIC}
\end{equation}
where $\mathcal{L}_\mathrm{max}$ is the maximum likelihood and $k$ is the number of free parameters.
The lower the value of AIC, the better the model. As we used \chisq\ minimization in our fitting, we
make use of the fact that $-2 \ln \mathcal{L}_\mathrm{max} = \chisq_\mathrm{min} + C$, where
$\chisq_\mathrm{min}$ is the best-fit value of \chisq, and $C$ is a constant independent of the
particular model being considered (as only differences in AIC are meaningful, we can ignore $C$).
For each foreground model, we calculated the AIC relative to that obtained with no foreground model,
\begin{equation}
  \Delta \mathrm{AIC} (\mbox{Model X}) = \mathrm{AIC} (\mbox{Model X}) - \mathrm{AIC} (\mbox{No f/g}).
\end{equation}
For the pn-only fits, the LB, C14-SWCX, and ACX-SWCX foreground models yield $\Delta \mathrm{AIC} =
-16.2$, $-8.2$, and $-11.6$, respectively. These differences in AIC amount to strong ($\Delta
\mathrm{AIC} < -5$) or decisive ($\Delta \mathrm{AIC} < -10$) evidence in favor of including a
foreground component in the model \citep{liddle07}.

\section{DISCUSSION AND CONCLUSIONS}
\label{sec:Discussion}

\subsection{Foreground Emission}
\label{subsec:DiscForeground}

The foreground emission toward G225 appears to be relatively faint in the \xmm\ band. Our spectral
analysis implies foreground 0.4--1.0~\kev\ surface brightnesses of 5.0 (2.3--8.8), 4.1 (2.4--8.2),
and 2.6 (0.6--4.4) $\times 10^{-13}~\flux\ \pdegsq$ for the LB, C14-SWCX, and ACX-SWCX foreground
models respectively (the values in parentheses are the 90\%\ confidence intervals).  In contrast,
the results of previous \xmm\ and \suzaku\ shadowing studies imply foreground 0.4--1.0~\kev\ surface
brightnesses of $(\mbox{7--18}) \times 10^{-13}~\flux\ \pdegsq$
\citep{lei09,gupta09b,smith07a,galeazzi07,henley07}. The highest of these is from a pair of
\xmm\ pointings on and off an unnamed dusty filament \citep{henley07}, which are now known to be
contaminated by stronger-than-typical SWCX emission \citep{koutroumpa07,henley08a}.

The faintness of the foreground emission limits the amount of physical information about the
foreground that we can extract from our observation of G225. For example, from the C14-SWCX model we
obtain only upper limits on the foreground \OVII\ \Kalpha\ and \OVIII\ \Lyalpha\ intensities, and so
we cannot constrain the solar wind \Oplus{8}/\Oplus{7} ion ratio using this model. The temperature
of the ACX-SWCX model can provide information on this ion ratio, albeit under the assumption of a
CIE ion distribution.  At the best-fit temperature of the ACX-SWCX component, $7.5 \times 10^5~\K$,
99\%\ of the oxygen is in the \Oplus{6} charge state (from ATOMDB v2.0.2;
\citealt{foster12}). Assuming CIE therefore results in a best-fit model from which there is
virtually no oxygen SWCX emission in the \xmm\ band (the SWCX emission from this model in the
\xmm\ band is mainly from \NVI\ \Kalpha\ and \CVI\ \Lybeta\ and \Lygamma).

We use the upper limit of the temperature of the ACX-SWCX component, $1.28 \times 10^6~\K$, to place
an upper limit of 0.006 on the solar wind \Oplus{8}/\Oplus{7} ratio (ATOMDB). This is significantly
less than the ratio expected for the slow solar wind (0.35; \citealt{schwadron00}), suggesting that,
during the \xmm\ observation, the portion of the G225 sight line in the heliosphere passed mainly
through fast solar wind (for which this ratio is nearly zero; \citealt{schwadron00}). This is a
somewhat surprising result, as the observation was taken only $\sim$9~months before the most recent
solar maximum (based on the sunspot number and the solar 1--8~\angstrom\ X-ray flux;
\citealt{winter14}), at which time the solar wind would be expected mostly to be slow
\citep{smith03}. Note that the upper limit on the ACX component's temperature (and hence on the
solar wind \Oplus{8}/\Oplus{7} ratio) is determined not just by the oxygen K lines, but also by
lower-energy lines from carbon and nitrogen, and so the low solar wind \Oplus{8}/\Oplus{7} ratio
could in principle be an artifact of our assuming the default \citet{anders89} abundances for the
ACX model. In practice, this appears not to be the case: if we adjust the abundances of carbon,
nitrogen, and neon relative to oxygen\footnote{The absolute abundances of these elements relative to
  hydrogen are not important here, as hydrogen does not emit in the \xmm\ band. The absolute
  abundances affect only the overall normalization of the ACX model.} so they match those expected
for the slow solar wind \citep[specifically, the average of the ``Max'' and ``Min'' values from
  their Table~1]{vonsteiger00} and refit, we find that the halo results are unaffected, and the
resulting upper limit on the solar wind \Oplus{8}/\Oplus{7} ratio is 0.010, still much lower than
the value expected for the slow solar wind. However, we note that the results for the ACX model
could be affected by the assumption of an ion distribution described by a single temperature.

It should also be noted that the sun was less active during the most recent maximum than during
previous maxima (e.g., the sunspot number and the solar 1--8~\angstrom\ flux at the most recent
maximum were approximately half the values at the 1990 maximum; \citealt{winter14}), which may have
affected the solar wind structure. Unfortunately, solar wind charge distribution data from the SWICS
instrument on board the \textit{Advanced Composition Explorer} (\ace) are unavailable for times
after August 2011,\footnote{http://www.srl.caltech.edu/ACE/ASC/level2/lvl2DATA\_SWICS-SWIMS.html}
whereas our observation was taken in February 2013. Therefore, we are unable to check if the solar
wind had an unusual ion composition prior to and during our observation.

\subsection{Halo Emission}
\label{subsec:DiscHalo}

The halo parameters derived from the G225 pn spectra are not sensitive to the particular foreground
model used in the analysis, although omitting the foreground component altogether does result in a
halo temperature that is $\sim$10\% lower. This insensitivity to the details of the foreground model
is likely due to the relative faintness of the foreground emission, noted above.  If the spectral
analysis carried out here were repeated on a shadowing observation with bright foreground emission,
we would expect to see some sensitivity of the halo parameters to the assumed form of the foreground
emission. We plan to test this in a future study. (Note that this will necessarily involve using
shadowing observations that consist of separate on- and off-shadow pointings, unlike the
single-pointing observation studied here).

G225 is included in the \citet{snowden00} catalog of X-ray shadows in the \rosat\ All-Sky Survey, as
shadow S2267M661.\footnote{This name is derived from the coordinates of the center of the region of
  the sky analyzed by \citet{snowden00}, rather than from the coordinates of the cloud.}  The
intrinsic 1/4~\kev\ halo count rate in the direction of G225 is $(947 \pm 178) \times
10^{-6}~\rassrate$. In contrast, our best fit halo models imply 1/4~\kev\ count rates of
$(\mbox{240--270}) \times 10^{-6}~\rassrate$.\footnote{These were calculated using v2.0.2 of APEC
  \citep{foster12}. If we instead use the \raymondsmith\ code, we obtain count rates $\sim$$30
  \times 10^{-6}~\rassrate$ higher.} This discrepancy implies that a $1T$ model cannot adequately
model the halo X-ray emission down to photon energies of $\sim$0.1~\kev, as was previously
demonstrated using \rosat\ All-Sky Survey data \citep{kuntz00}. In order to obtain a reasonable
model of the 1/4~\kev\ emission, a $\sim$$1 \times 10^6~\K$ component must be added to the halo
model---since such a component would contribute to the halo's \OVII\ emission, its inclusion would
affect the best-fit temperature of the $\sim$$2 \times 10^6~\K$ component of our current spectral
model. Even such a two-temperature model is likely an approximation of the halo's true temperature
structure, as there may be a continuum of temperatures in the halo \citep{shelton07,lei09}. However,
the $1T$ halo model used here is still useful for characterizing the emission within the
0.4--5.0~\kev\ \xmm\ band, and the results obtained from such halo models can still be used to test
models of the hot halo gas, provided such models' emission predictions are characterized in the same
way as the observed emission \citep{henley10b}.

\begin{figure}
\centering
\plotone{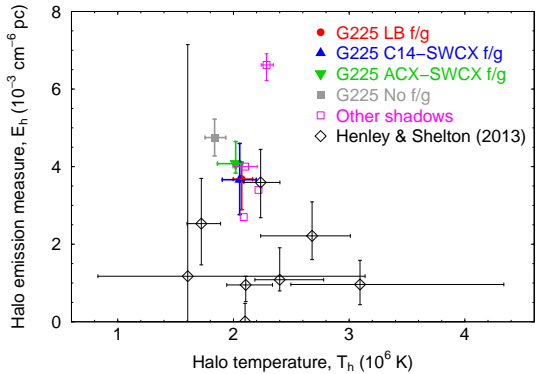}
\caption{Comparison of our halo measurements with those from previous studies. The solid symbols
  show our pn-only results from Figure~\ref{fig:HaloResults}.  The magenta squares show the results
  from previous \xmm\ or \suzaku\ shadowing studies: from top to bottom, a \suzaku\ study of an
  unnamed dusty filament (\citealt{lei09}; note that this result has been rescaled---see text for
  details), a \suzaku\ study of MBM~12 \citep{smith07a}, an \xmm\ study of MBM~20
  \citep{galeazzi07}, and a \suzaku\ study of MBM~20 \citep{gupta09b}. The black diamonds show
  results from the \citet{henley13} \xmm\ survey of the halo, for sight lines within 15\degr\ of
  G225.
  \label{fig:Comparison}}
\end{figure}

Figure~\ref{fig:Comparison} compares our measurements with those from previous \xmm\ and
\suzaku\ shadowing studies. In these studies, the halo emission was characterized with a single
X-ray temperature. The \citet{lei09} result was obtained using a different abundance table from the
other studies (\citealt{wilms00} versus \citealt{anders89}\footnote{Note that \citet{gupta09b} do
  not explicitly state which abundance table they used for their plasma emission components.}).  The
halo emission is dominated in the \xmm/\suzaku\ band by oxygen \Kalpha\ emission; for a given
temperature, the intensity of this emission is proportional to $\int \Ne \nO dl = \EMh A_\mathrm{O}
/ 1.2$, where \Ne\ and \nO\ are the halo electron and oxygen number densities, respectively, $\EMh
\equiv \int \Ne^2dl$ is the halo emission measure, and $A_\mathrm{O}$ is the halo oxygen
abundance. Hence, the best-fit halo emission measure is approximately inversely proportional to the
assumed value of $A_\mathrm{O}$. Therefore, in order to allow a fair comparison with the other
results, we have multiplied the \citet{lei09} emission measure by
$A_\mathrm{O}(\mbox{\citeauthor{wilms00}}) / A_\mathrm{O}(\mbox{\citeauthor{anders89}}) = 0.576$.

In general, our results are in good agreement with those from previous \xmm\ and \suzaku\ shadowing
studies. This agreement implies that the fact that these other studies consisted of two separate
pointings, which could potentially have had different foreground brightnesses (see Introduction),
did not adversely affect the halo results. However, as noted above, the halo results derived from
these other studies may be sensitive to the assumed foreground model.

Figure~\ref{fig:Comparison} also compares our measurements with results for nearby sight lines in
the \citet{henley13} \xmm\ survey of the halo (within 15\degr\ of G225). In this survey, the
foreground model was based on results from the previously mentioned \citet{snowden00} shadow
catalog, extrapolated from the 1/4~\kev\ \rosat\ band to the 0.4--5.0~\kev\ \xmm\ band. The
\citet{henley13} emission measures shown in Figure~\ref{fig:Comparison} are typically smaller than
that obtained from G225. One might therefore conclude that there is a systematic error in the
\citeauthor{henley13} emission measures, possibly due to the assumed foreground model. However, the
\citeauthor{henley13} result that is closest to the G225 results in Figure~\ref{fig:Comparison}
(obs.~0302500101, at $(\Th,\EMh) = (2.2 \times 10^6~\K, 3.6 \times 10^{-3}~\emismeas)$) is also the
closest sight line to G225 on the sky (angular separation = $4\fdg8$). Hence, it may simply be that
the halo within a few degrees of G225 is somewhat brighter than its surroundings. (Note that this
does not preclude the possibility that other, more distant regions of the halo are also bright---the
other shadows whose results are plotted in Figure~\ref{fig:Comparison} are $\sim$30--50\degr\ from
G225.) Furthermore, the agreement between the G225 measurements and the measurement from the nearest
\citet{henley13} sight line supports the conclusion that the \citeauthor{henley13} results are well
calibrated and not subject to systematic errors. Such a conclusion is important for when we use the
\citeauthor{henley13} measurements to test models of the halo X-ray emission \citep{henley14c}.

\acknowledgements
We thank the anonymous referee, whose comments have helped improve this paper.
This research is based on observations obtained with \xmm, an ESA science mission with instruments
and contributions directly funded by ESA Member States and NASA.
We acknowledge use of the R software package \citep{R}.
This research was funded by NASA grant NNX13AF69G, awarded through the Astrophysics Data Analysis
Program, and partially supported by NASA grant NNX09AC46G.

\bibliography{references}

\end{document}